\documentclass[twocolumn,pra,aps,superscriptaddress,showpacs]{revtex4-1}
\usepackage{amsmath}
\usepackage{amssymb}
\usepackage{amsfonts}
\usepackage{amssymb}
\usepackage{stackrel}
\usepackage[dvips]{graphicx}
\usepackage{subfigure} 
\usepackage{dcolumn}
\usepackage{txfonts}
\usepackage{bm}
\usepackage{makeidx}
\usepackage{color}
\usepackage{mathtools}
\usepackage{threeparttable}
\usepackage{upgreek}
\usepackage[colorlinks,linkcolor=blue,anchorcolor=blue,citecolor=blue,urlcolor=blue]{hyperref}

\begin{document}
\title{Sensitivity of Displacement Detection for a Particle Levitated in the Doughnut Beam}

\author{Lei-Ming Zhou}
\address{Beijing Computational Science Research Center, Beijing 100193, China}

\author{Ke-Wen Xiao}
\address{Beijing Computational Science Research Center, Beijing 100193, China}

\author{Zhang-Qi Yin}
\address{Center for Quantum Information, Institute of Interdisciplinary Information
Sciences, Tsinghua University, Beijing 100084, China}

\author{Jun Chen}
\address{Institute of Theoretical Physics and Collaborative Innovation Center
of Extreme Optics, Shanxi University, Shanxi, China}

\author{Nan Zhao}
\email{nzhao@csrc.ac.cn}
\address{Beijing Computational Science Research Center, Beijing 100193, China}

\date{\textcolor{blue}{\today }}

\begin{abstract}
Displacement detection of a sphere particle in focused laser beams with quadrant
photodetector (QPD) provides a fast and high precision way to determine
the particle location. In contrast to the traditional Gaussian beams,  the
sensitivity of displacement detection using various doughnut beams are investigated. The sensitivity
improvement for large sphere particles along the longitudinal direction is reported. With appropriate vortex charge $l$ of the doughnut beams, they can outperform the Gaussian beam to get more than one order higher sensitivity and thus have potential applications in various high precision measurement.  By using the levitating doughnut beam itself to detect the particle displacement, the result will also 
facilitate the recent proposal of
levitating a particle in doughnut beams to suppress the light absorption.
\end{abstract}

\maketitle

\textbf{\textcolor{blue}{\emph{Introduction}}}\emph{\label{sec:Introduction}}
The displacement of the levitated particle is usually measured by
the interferometry method with quadrant photodetector (QPD) in the
back focal plane \cite{DenkAO1990Optical,SvobodaNature1993direct,FinerNature1994single,GittesOL1998Interference,PralleMRT1999Three,RohrbachJAP2002Three,RohrbachOL2003Three,VolpeJAP2007Backscattering}. With high sensitivity and high bandwidth, it is widely used in high precision displacement measurement \cite{DenkAO1990Optical}, weak force measurement \cite{SvobodaNature1993direct,FinerNature1994single}, photon force microscope \cite{FlorinJSB1997photonic}, optical nanoprobing \cite{HerreraPRL2010Optical}
and even surface imaging \cite{FriedrichNNano2015Surface}. 
Especially, with the sensitivity as high as 3 $\rm{fm/\sqrt{Hz}}$ \cite{LiPhD2012fundamental}, it can measure the instantaneous velocity of brownian particle \cite{LiTCScience2010,LiPhD2012fundamental} and provides a key tool to investigate the dynamics of the
particle in various physical systems \cite{AritaPRA2016Orbital,jiangNPhysics2010nonlinear} including the optomechanical system \cite{Neukirch2015NPhotonicsMulti,HoangNCom2016}. 

Particle levitated by laser beam absorbs light and meets heating problem \cite{Neukirch2015NPhotonicsMulti,HoangNCom2016}. It is proposed to reduce heating of strong absorptive particle by designing a core-shell structure of the particle and trapping it in
the doughnut-shaped beams \cite{ZhouLPR2017optical}. The proposal of the
using doughnut beams show excellent tolerance of the heat absorption of
the particle and keeps the high quality factor of the mechanical oscillation. When the laser beam is changing from Gaussian beam to doughnut beams,
it is significant to know how the doughnut beams affect  the sensitivity in detecting the particle displacement.

The interferometry method with QPD has been investigated extensively using Gaussian beam \cite{DenkAO1990Optical,SvobodaNature1993direct,FinerNature1994single,GittesOL1998Interference,PralleMRT1999Three,RohrbachJAP2002Three,RohrbachOL2003Three,VolpeJAP2007Backscattering,FriedrichNNano2015Surface}.
However, few papers have investigated the sensitivity of displacement detection
of a particle in doughnut beams \cite{vandeNesOE2007,GarbinCLEO2008mie,GarbinNJP2009Mie}.
Nes et. al. \cite{vandeNesOE2007} has investigated the scattering
of a sphere particle in LG beam and shown the response signal of QPD for certain LG
beams. Garbin et. al. \cite{GarbinNJP2009Mie}
have investigated the signal of QPD for LG beams with different sign
of vortex charge $l$. It has shown that the result can be used to
distinguish the vortex charge of the beam. Here, we give a detail
and systematical investigation of interferometry method using various
doughnut beams. Especially, the displacement detection with the
large sphere particle size has been included. The result shows that the interferometry method is still efficient for doughnut beams. More importantly, the sensitivity along the longitudinal direction is dramatically improved for large sphere particles. By choosing appropriate doughnut beams, they can outperform the Gaussian beam with more than one order.

\begin{figure}[tb]
\begin{centering}
\includegraphics[width=8cm]{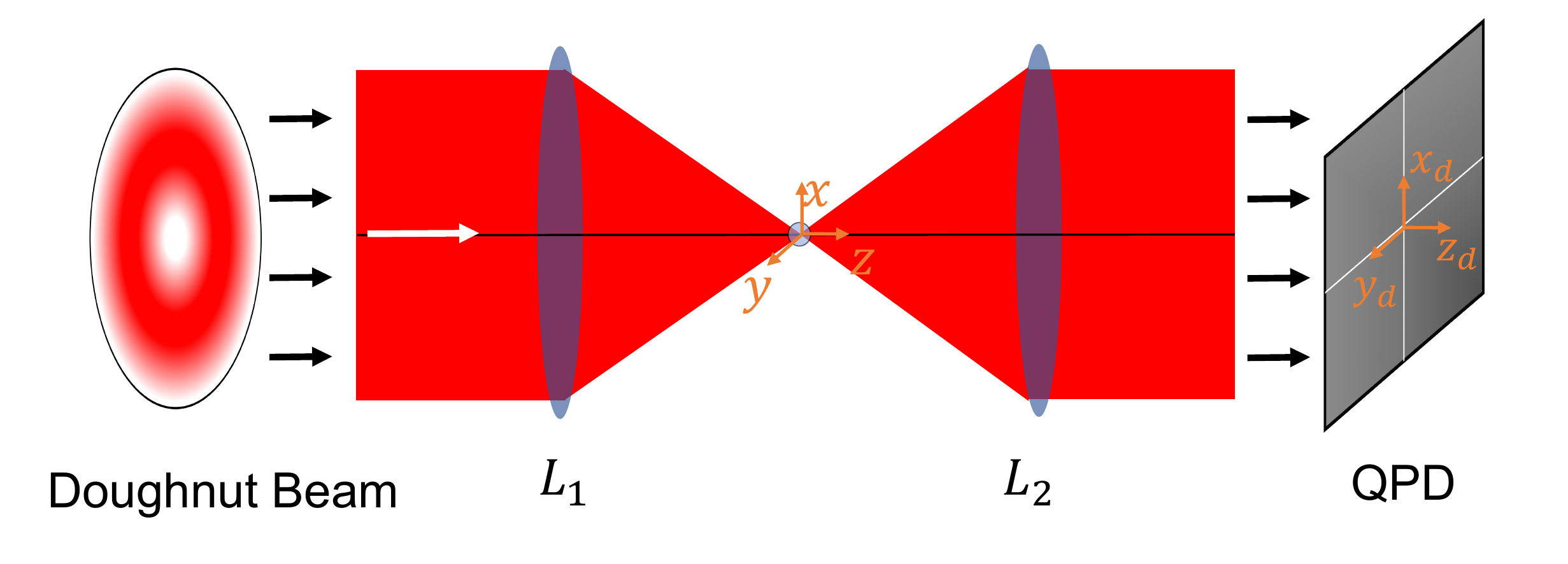}
\end{centering}
\caption{The schematic of the displacement detection setup in an optical levitating system (consisted of an optical trap and a trapped particle). Incident
beams here include various doughnut beams. The field after the particle
scattering is collected by lens ${\rm L}_{2}$ to the quadrant photodetector
(QPD) to get the signal.\label{fig:trap_skematic}}
\end{figure}

\textbf{\textcolor{blue}{\emph{The system structure and method}}}\textcolor{blue}{\label{sec:The-structure-of-OT}}
The system we considered is shown in Fig.~\ref{fig:trap_skematic}. The incident
laser beam is focused by a high Numerical Aperture (NA) objective
lens ${\rm L}_{1}$ to trap a sphere particle near the focus in vacuum. The scattering
beam field, along with the incident beam field is collected by
the second lens ${\rm L}_{2}$ to the QPD as the detection signal.
In the previous literatures \cite{GittesOL1998Interference,PralleMRT1999Three,RohrbachJAP2002Three,RohrbachOL2003Three,VolpeJAP2007Backscattering}, the incident beam is typically Gaussian beam.
Here we focused on various doughnut beams. The system is described
with the following parameters: the power of incident beam $P$, the wavelength
of the beam in vacuum $\lambda_{0}$, the polarization of incident
beam $\mathbf{e}$, the numerical aperture of the object lens ${\rm NA}$, the filling factor of the incident beam  $f_{01}$ and the radius of sphere particle $a$. Unless stated otherwise, we assume $P=100\ {\rm mW}$, $\lambda_{0}=1064\ {\rm nm}$, ${\rm NA}=0.95$, $\mathbf{e}=\mathbf{\hat{\mathbf{x}}}$ and $f_{01}=1.0$ in this work. 

We apply the generalized Lorentz-Mie Theory (GLMT) to simulate the
electromagnetic field scattering by the particle. The GLMT has been widely
used in literatures \cite{Nieminen2007Optical,ChenPRE2009}, and provides a powerful and convenient tool to calculated the scattering field.
In this method, the incident and scattering beam are described by
the partial wave expansion coefficients in the bases of vector spherical
wave functions. The two sets of coefficients are linked by the $T$-Matrix
of the particle which is independent of the incident beam. The strongly focused
incident beam cannot be described by the expression of paraxial beam, thus we
described the beam by the generalized vector Debye integral theory \cite{ChenPRE2009}.

The incident and scattered electromagnetic fields are collected by the lens ${\rm L}_{2}$. The QPD then outputs the response signal based on the interference distribution of the light intensity. The responses signals of the QPD are \cite{JonesBook2015optical}
\begin{subequations}
\begin{eqnarray}
S_{x} & = & \underset{x_{d}>0}{\iint}I(x_{d},y_{d})dx_{d}dy_{d}-\underset{x_{d}<0}{\iint}I(x_{d},y_{d})dx_{d}dy_{d},\label{eq:singalSx} \\
S_{y} & = & \underset{y_{d}>0}{\iint}I(x_{d},y_{d})dx_{d}dy_{d}-\underset{y_{d}<0}{\iint}I(x_{d},y_{d})dx_{d}dy_{d},\label{eq:singalSy}\\
S_{z} & = & \iint I(x_{d},y_{d})dx_{d}dy_{d}.\label{eq:singalSz}
\end{eqnarray}
\end{subequations}
Among Eqs.~(\ref{eq:singalSx}-\ref{eq:singalSz}), the subscripts $x,y,z$ denote the QPD's three different signals respectively. Usually, $S_{x}$ is chosen to measure the particle displacement for a particle moving along the $x$ direction for example, because the signal $S_x$ changes most dramatically at the same time. So are the cases of $S_{y}$ and $S_{z}$. Since both $S_{x}$ and $S_{y}$ denote the signals in the transverse direction and have similar behaviors, we mainly show the results of $S_x$ and $S_{z}$ in the investigation below without loss of generality. The light intensity on the detector is $I(x_{d},y_{d})=\frac{c\varepsilon_{d}}{2n_{d}}|\mathbf{E}(x_{d},y_{d})|^{2}$, where the permittivity
and refractive index of medium before the detector are $\varepsilon_{d}$ and $n_{d}$ respectively. The velocity
of light in vacuum is denoted as $c$. The field $\mathbf{E}(x_{d},y_{d})$ is the electric field
on the detector. It can be written as
%\begin{equation}
$\mathbf{E}(x_{d},y_{d})={\mathbf{E}_{\rm inc}}(x_{d},y_{d})+{\mathbf{E}_{\rm scat}}(x_{d},y_{d})$, %\label{eq:Esuperposition}
%\end{equation}
% In Eq.~(\ref{eq:Esuperposition}),  
where ${\mathbf{E}_{\rm inc}}(x_{d},y_{d})$ comes from the incident beam
which can be expressed analytically for the confocal system here; % (More details in the Supplementary Information);
and ${\mathbf{E}_{\rm scat}}(x_{d},y_{d})$ comes from the scattering filed, which is calculated
through the $T$-matrix method.

\textbf{\textcolor{blue}{\emph{Result}}}\textcolor{blue}{\label{sec:Result}}
In this part, we show the numerical result of displacement detection sensitivity for a particle
in Gaussian beam, and various doughnut beams. The definition of doughnut beams including LG beam with different vortex charge
$l$, radially polarized and azimuthally polarized beam follows the convention in Ref. \cite{JonesBook2015optical}.

\begin{figure}[t]
\begin{centering}
\includegraphics[width=8cm]{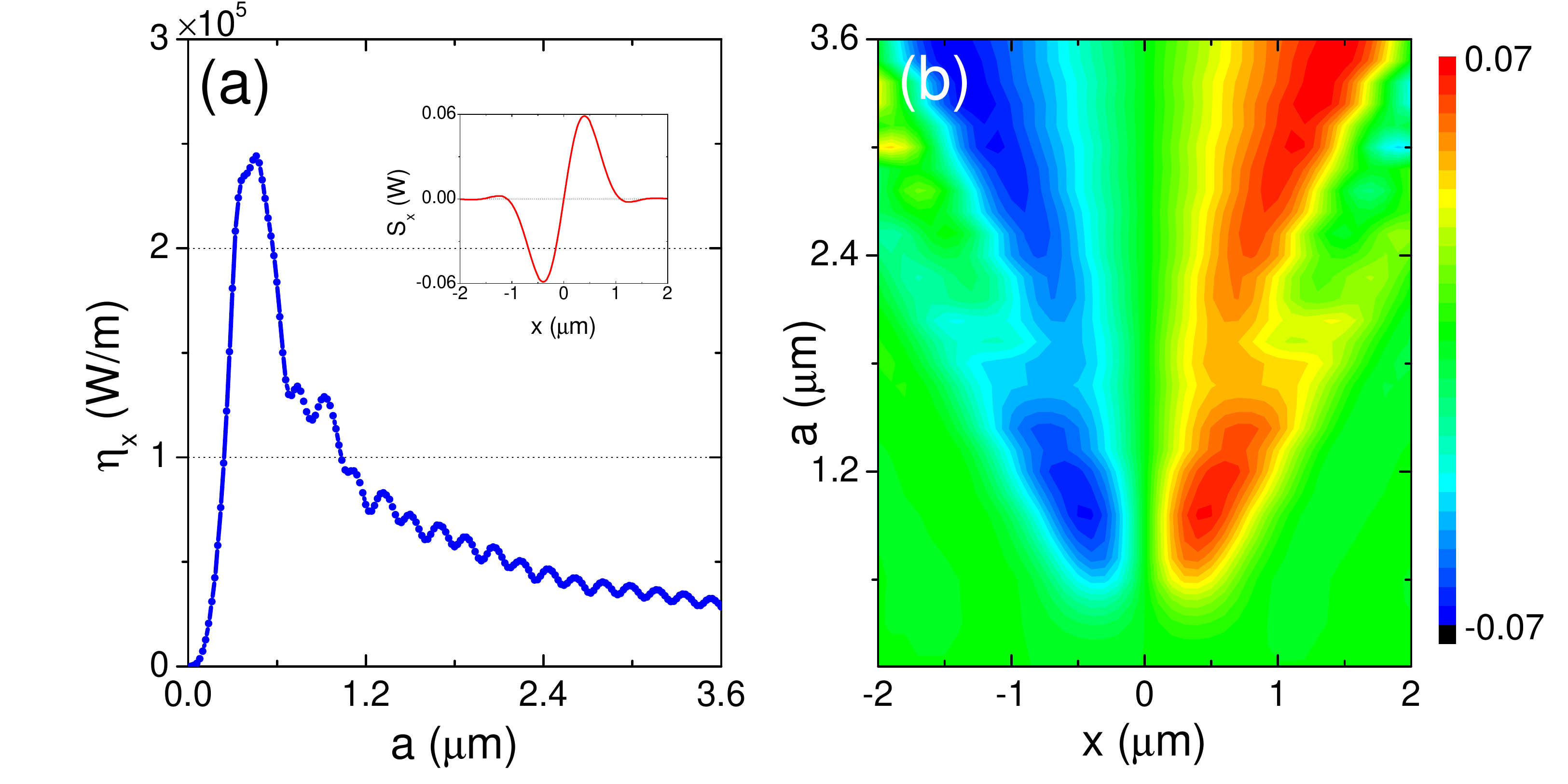}
\end{centering}
\caption{(a) The sensitivity of displacement detection using QPD for a silica
sphere particle trapped in the $\hat{\mathbf{x}}$ direction linearly polarized
Gaussian beam. The inset: a typical response of the QPD for a sphere particle ($a=400\ {\rm nm}$)
moving along the x axis in the Gaussian beam. (b) The response signal
of the QPD for particles with various radii in the $\hat{\mathbf{x}}$
direction linearly polarized Gaussian beam.\label{fig:S_R_Gaussian}}
\end{figure}

\begin{figure}[tb]
\begin{centering}
\includegraphics[width=8cm]{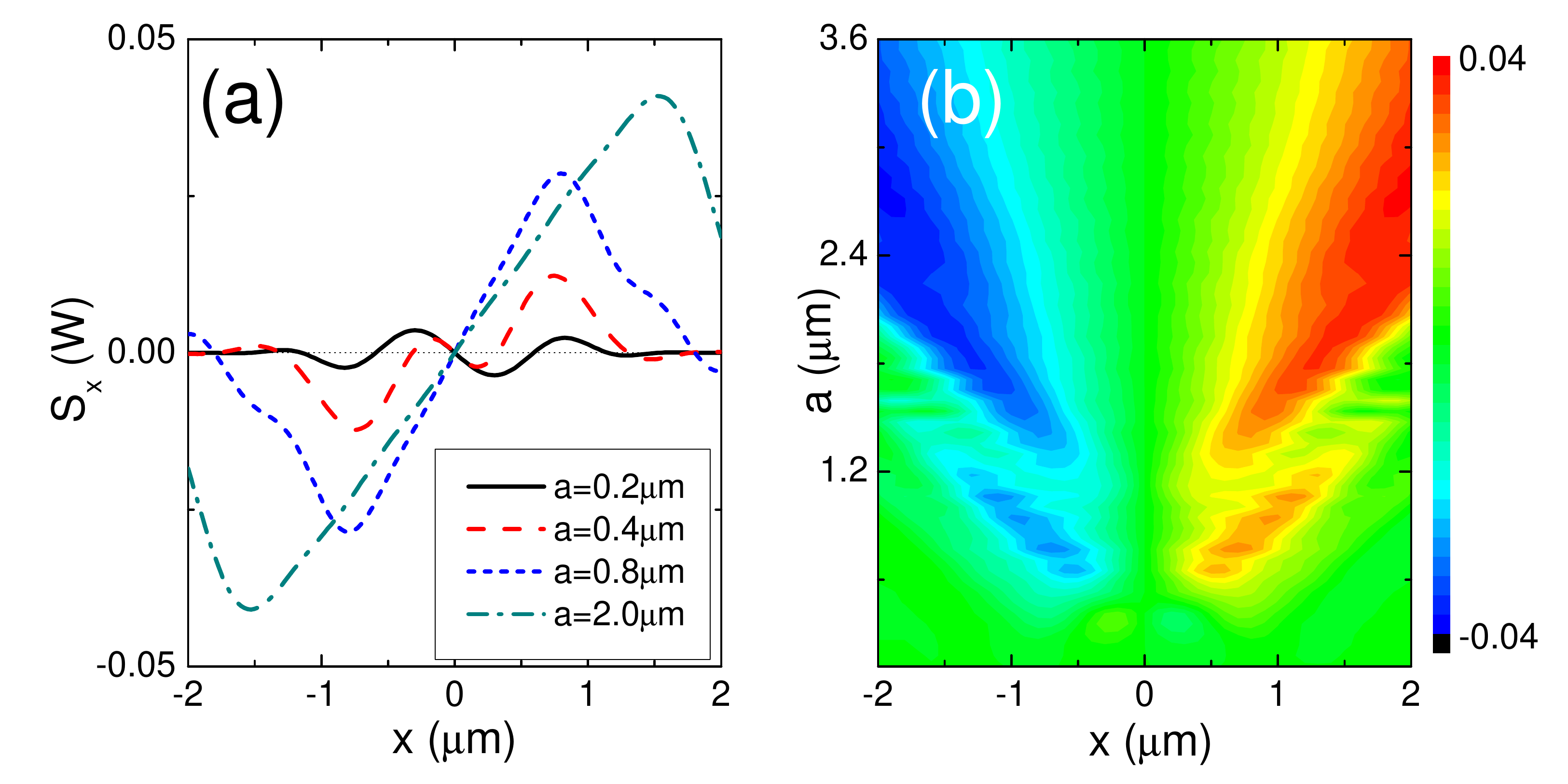}
\end{centering}
\caption{(a) The typical response of the QPD for a silica particle particle moving
along the x axis in the $\hat{\mathbf{x}}$ direction linearly polarized
${\rm LG}_{01}$ beam. Cases with some representative particle radii
have been shown. The slope near the beam focus (i.e., the sensitivity)
could be either positive or negative. (b) The response signal of the
QPD for particles with various radii in the $\hat{\mathbf{x}}$ direction
linearly polarized ${\rm LG}_{01}$ beam.\label{fig:QPDsignalLG}}
\end{figure}

\textcolor{blue}{\emph{Sensitivity and Range of a sphere particle in Gaussian beam}}
Figure \ref{fig:S_R_Gaussian}(a) shows the result of the detection sensitivity using Gaussian beam, for a silica sphere particle moving along $\hat{\mathbf{x}}$ direction in vacuum. Especially, the case for the sphere particle with
large size (i.e., much larger than the wavelength) is included. In the inset of Fig.~\ref{fig:S_R_Gaussian}(a), the response signal of the QPD for a sphere with radius $a=0.4\ \upmu {\rm m}$ is shown. The sensitivity is defined as
\begin{equation}
\left. \eta_{i}=\frac{dS_{i}}{dr_{i}} \right|_{\mathbf{r}=\mathbf{r}_0},\label{eq:Sensitivity}
\end{equation}
where $i=x, y, z$ denote the sensitivity along different displacement
direction and ${\mathbf{r}_0}$ is the location of the particle. Here we focus on ${\mathbf{r}_0}=0$ in the work because the sensitivity usually reaches it best at the axis origin.  As shown in Fig.~\ref{fig:S_R_Gaussian}(a), the sensitivity $\eta$ increase with the increasing
sphere particle radius $a$ and reaches the maximum when $a\approx0.4\ \upmu {\rm m}$.
At this point, the particle size is comparable with the beam waist. With larger radius $a$, the sensitivity decreases and shows shallow modulations. The shallow
modulations in the sensitivity curve when $a>0.4\ \upmu {\rm m}$
are caused by the Mie resonance which changes the scattering field
distribution.

\begin{figure}[t]
\begin{centering}
\includegraphics[width=8.5cm]{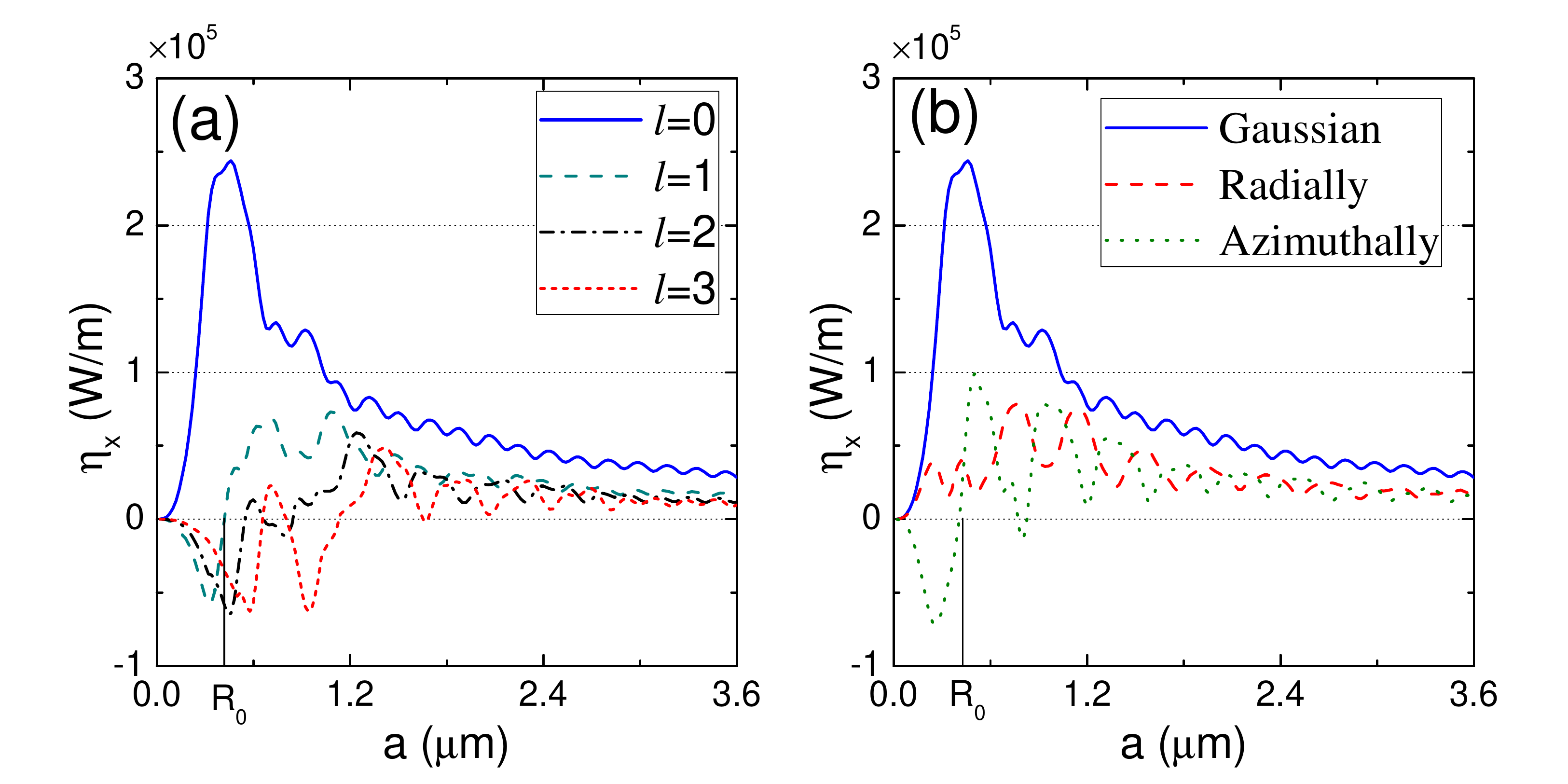}
\end{centering}
\caption{(a) The sensitivity of the transverse displacement detection for different particle
radii in the $\hat{\mathbf{x}}$ direction linearly polarized ${\rm LG}_{01}$,
${\rm LG}_{02}$ and ${\rm LG}_{03}$ beam and Gaussian beam (i.e., $l=0$).  (b) The same as Fig.~(a) for radially and azimuthally polarized beams.
\label{fig:S_R_LG}}
\end{figure}

\textcolor{blue}{\emph{Sensitivity of a particle in doughnut beams}} 
The QPD
signal of a sphere particle in LG beams is quite different from that
in Gaussian beam. As shown in Fig.~\ref{fig:QPDsignalLG}(a), taking the ${\rm LG}_{01}$ beam as an example,
there are more maximums and minimums of the signal $S_{x}$ when
the particle location changes. The slope of $S_{x}$ about $x$ at the coordinate origin, which
is defined as sensitivity, could be either positive or negative depending on the radius $a$. 
Figure~\ref{fig:S_R_LG}(a) shows the result of the sensitivity $\eta_{x}$ for ${\rm LG}_{0l}$ beams with $l=1,2,3$. The sensitivity for Gaussian beam (i.e., $l=0$) is also shown for comparison. For particles with size $a$ much larger than the beam waist size, the sensitivity of LG beams show the same behavior with that of Gaussian beam. However, for $a$ smaller than the beam waist size, the sensitivity  of LG beams (taking ${\rm LG}_{01}$ beam as an example) will change from negative to positive at certain size $R_{0}$. The displacement
direction can be positive or negative when the response signal is
positive, depending on whether the size of the particle exceeds $R_{0}$. This is different
from that in Gaussian beam (blue solid line), which is always positive.

The different behaviors here are caused by the doughnut shape
of intensity distribution of LG beams. The scattering field will show
totally different distribution depending on two factors. One is whether the
particle is inside or outside the bright rings of LG beams. The other
one is whether the particle radius is larger than the size of the
bright rings. To show this more clearly, the QPD signal $S_{x}$ are
plotted with various sphere radii and locations in Fig.~\ref{fig:QPDsignalLG}(b)
and Fig.~\ref{fig:S_R_Gaussian}(b) for linearly polarized
${\rm LG}_{01}$ beam and Gaussian beam, respectively. For the
${\rm LG}_{01}$ beam, the QPD signal shows different behaviors when the particle size is smaller
than the beam waist size. There are more lobes which affect the sensitivity.

For Gaussian beam, the sensitivity increases first and then decreases with the increasing size of the sphere particle, and there is a tradeoff between
the sensitivity and linear range of displacement detection. It is the same for LG beams only when the particle size is larger than the beam waist size. For smaller particles in LG beams, as the sensitivity sign depends on the particle size and changes dramatically near $R_{0}$, it could be used to measure the size of particle with high accuracy. The sensitivity for radially and azimuthally polarized beams is shown in Fig.~\ref{fig:S_R_LG}(b). The azimuthally polarized beam shows same behaviors as LG beams. The radially polarized beam shows similar behavior as the linearly polarized Gaussian beam, except the large modulations when the particle size is smaller than the beam waist size. 

\begin{figure}[t]
\begin{centering}
\includegraphics[width=8.5cm]{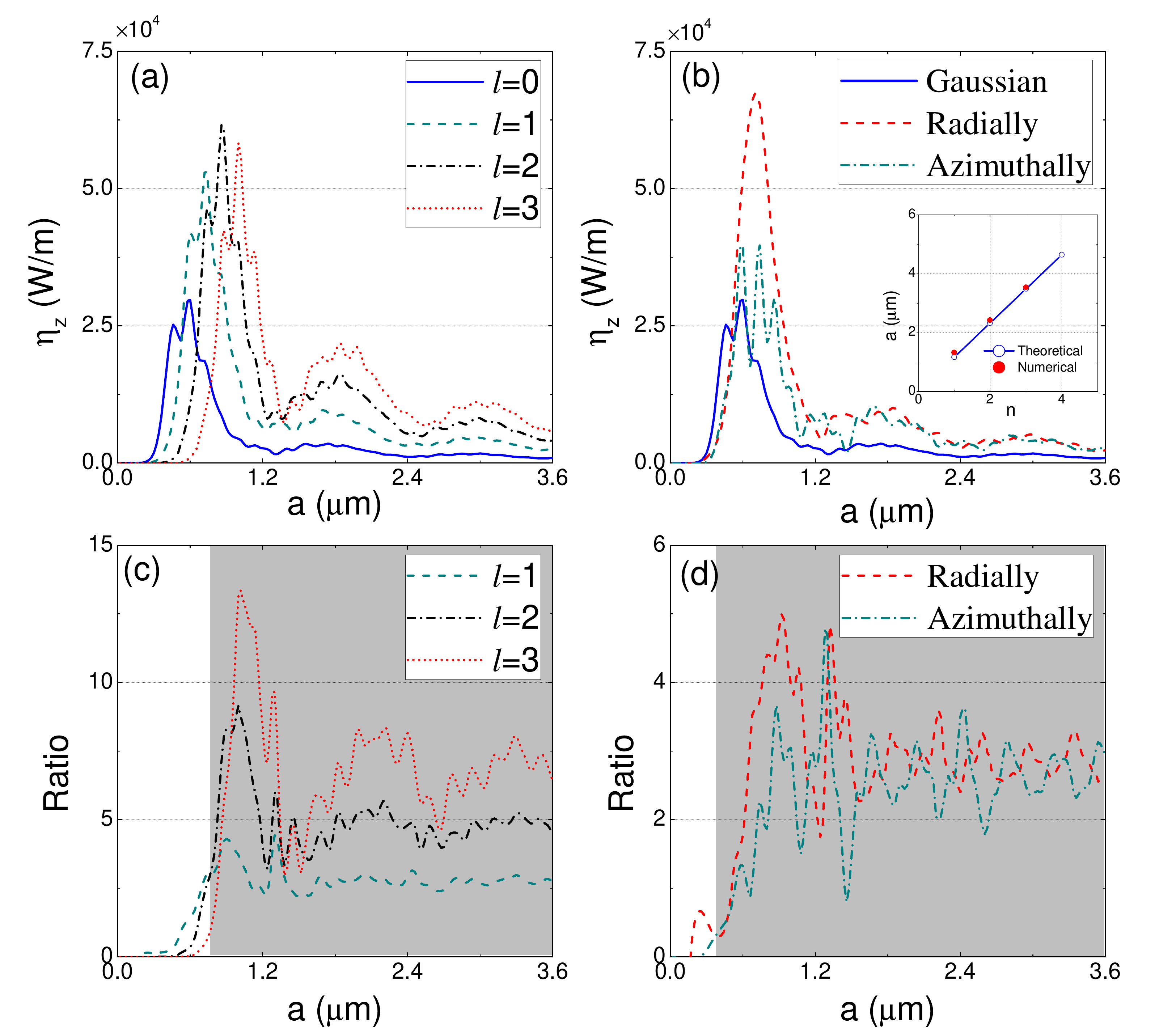}
\end{centering}
\caption{(a) The sensitivity $\eta_{z}$ of the longitudinal displacement detection, for different particle radii $a$ in the $\hat{\mathbf{x}}$ direction linearly polarized ${\rm LG}_{01}$,
${\rm LG}_{02}$ and ${\rm LG}_{03}$ beam and Gaussian beam (i.e., $l=0$).  (b) The same as Fig.~\ref{fig:S_R_LG_z}(a) for radially and azimuthally polarized beams. (c) The sensitivity improvement ratio $\xi$ for LG beams. (d) The same as Fig.~\ref{fig:S_R_LG_z}(c) for radially and azimuthally polarized beams. The region is colored grey in Fig.~\ref{fig:S_R_LG_z}(c) and Fig.~\ref{fig:S_R_LG_z}(d) where the particle can be trapped stably by $\rm{LG_{03}}$ beam and azimuthally polarized beam respectively. 
\label{fig:S_R_LG_z}}
\end{figure}

The sensitivity along longitudinal direction $\eta_{z}$ using various beams is shown in Fig.~\ref{fig:S_R_LG_z}(a) and Fig.~\ref{fig:S_R_LG_z}(b). Similar to the case for transverse displacement, the sensitivity increases first and then decreases with the increasing size of the sphere particle. 
With larger radius $a$, the sensitivity shows a large period
oscillation along with shallow modulations. The shallow
modulations is caused by the Mie resonance as in the transverse case. The large period oscillation is induced by the
interference between the incident beam propagating directly to the
detector and the scattering beam passing through the sphere particle to the
detector. The destruction interference point is then given by 
\begin{equation}
2ak_{0}(n_{m}-1)=2n\pi,\label{eq:Distruction Point}
\end{equation}
where $n$ are positive integers, $n_{m}=1.458$ is the refractive
index of the silica sphere particle and $k_{0}$ is the vacuum wavevector.
To verify this, minimum points for the case of Gaussian beam in Fig.~\ref{fig:S_R_LG_z}(b)
are plotted in its inset to
compare with the theoretical result given by Eq.~(\ref{eq:Distruction Point}).
Thus we also get the oscillation period 
\begin{equation}
\lambda_L=\frac{\lambda_0}{2(n_m-1)},\label{eq:lambdaL }
\end{equation}
and $\lambda_L=1.16\ \upmu {\rm m}$ here. The existence of these destruction
interference point has a disadvantage in the displacement detection for sphere particles with size $a=n\lambda_L$.

At the same time, it is noticed that the sensitivity $\eta_{z}$ in doughnut beam is higher than that in Gaussian beam, when the size of the particle is comparable or lager than the beam waist. We define the sensitivity improvement ratio as
\begin{equation}
\xi_{l}=\eta_{\text{z},l}/\eta_{\text{z,Gaussian}},\label{eq:Ratio}
\end{equation}
where $l=1,2,3$ denotes different LG beams. The result is shown in Fig.~\ref{fig:S_R_LG_z}(c). The improvement ratio of sensitivity can be more than one order. Especially, the grey region in Fig.~\ref{fig:S_R_LG_z}(c) shows where the particle could be trapped stably by LG beams (taking the ${\rm LG}_{03}$ beam as an example) \cite{ZhouLPR2017optical}. The region of improved sensitivity (i.e., $\xi>1$) falls just into the grey region, thus the LG beam provides higher sensitivity in LG beam levitated systems. The sensitivity improvement ratio for radially and azimuthally polarized beams is shown in Fig.~\ref{fig:S_R_LG_z}(d) and it has the same improvement behavior as LG beams.

\textbf{\textcolor{blue}{\emph{Discussion}\label{sec:Discussion}}} 
It is significant to know the best sensitivity we could get for a particle with
certain size. The systematically result for Gaussian beam
and various doughnut beams here will facilitate us to choose the proper
beam to get optimal sensitivity. Generally speaking, Gaussian beam
will get better sensitivity for transverse displacement detection. However, doughnut beams can have one order higher sensitivity for longitude displacement detection, when the particle has the size exceeds the beam waist.

As a conclusion, the sensitivity of displacement detection of a particle in various doughnut beams are studied. We
pay attention especially to the case of large particle size. The result for doughnut beams provides us the ability to
choose proper beam to get best sensitivity.  By using the levitating doughnut beam itself to detect the particle displacement, it will also benefit the recent proposal of levitating a particle using doughnut beams to suppress the light absorption.

Note: In the last version of this manuscript on arXiv, there was a mistake: the incident beam hasn't been normalized to the power $P=100\ {\rm mW}$ while the scattering beam has. So, as the incident beam has the power of only several percent of the scattering beam power, the last version showed approximately the signal behaviors of scattering beams.

\textbf{\textcolor{blue}{\emph{Acknowledgement}}}
We thank Prof. Zhi-Fang Lin and Dr. Wen-Zhao Zhang for the helpful discussion. This work is supported by Science Challenge Project No. TZ2018003, National Basic Research Program of China No. 2014CB848700, No. 2016YFA0301201, NSFC No. 11534002, NSAF U1730449 and NSAF No. U1530401. Z.-Q. Y. is supported by NSFC Grant 61771278, 61435007, and the Joint Fund of the Ministry of Education of China (6141A02011604). J.C. is supported by NSFC No. 11674204.

\appendix
\section{Field on the detector}

\begin{figure*}
\centering{}\includegraphics[width=10cm]{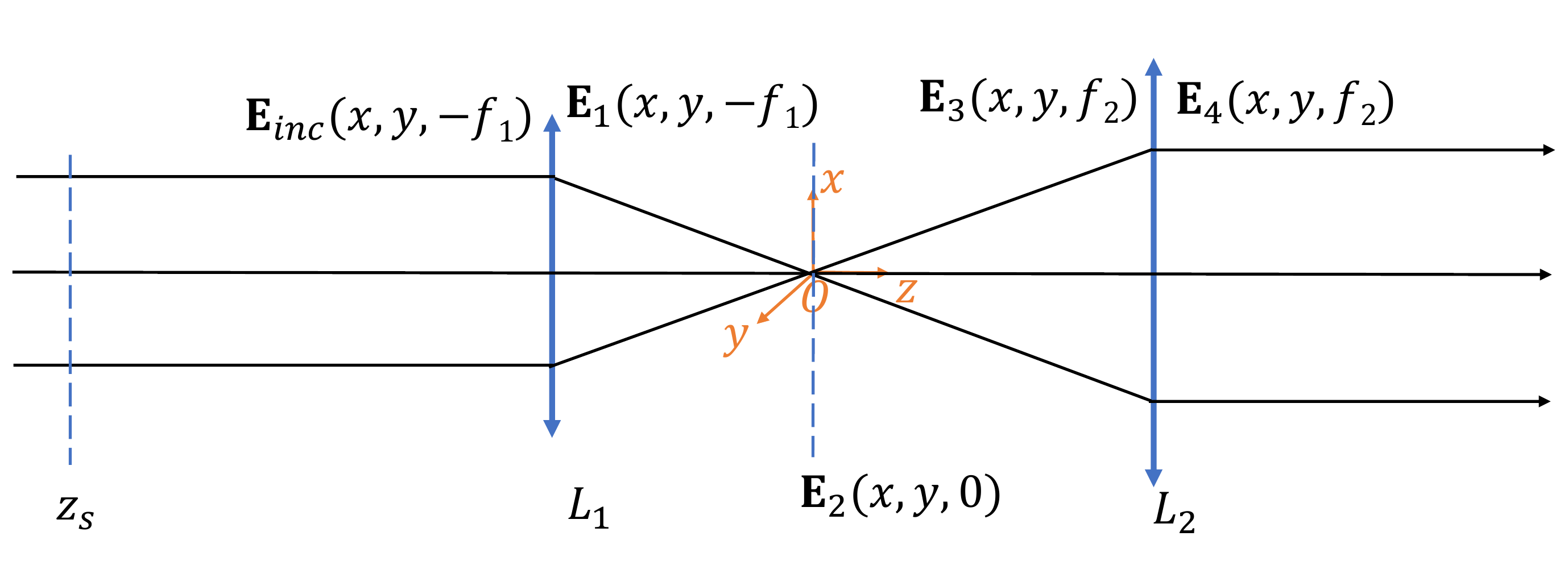}
\caption{Theoretical model of the detection system. Here $\mathbf{E}_{{\rm inc}}(x,y,-f_{1})$
and $\mathbf{E}_{1}(x,y,-f_{1})$ are used to denote the field on
the left and right side of the focusing lens $L_{1}$. Field $\mathbf{E}_{3}(x,y,f_{2})$
and $\mathbf{E}_{4}(x,y,f_{2})$ are used to denote the field on the
left and right side of the collecting lens (condensor) $L_{2}$. And
$\mathbf{E}_{2}(x,y,0)$ denotes the field on the $z=0$ plane. 
\label{fig:TheoreticalModel}}
\end{figure*}

For the detection system described in the main text, it can be simplified as a confocal system as show in Fig.~\ref{fig:TheoreticalModel}.
The incident field ${\mathbf{E}_{\rm inc}}(x_{d},y_{d})$ on the detector is denoted as $\mathbf{E}_{4}(x,y,f_{2})$ here. It  can be expressed
by the incident filed $\mathbf{E}_{{\rm inc}}(x,y,-f_{1})$ as
\begin{equation}
\mathbf{E}_{4}(x_{2},y_{2},f_{2})=-e^{ik(f_{2}+f_{1})}e^{il\phi}\mathbf{E}_{{\rm inc}}(-\frac{x_{2}}{M},-\frac{y_{2}}{M},-f_{1}).\label{eq:Field_detector_incident}
\end{equation}
To make our convention clear, we will show the procedures to get this relation here.

\subsection{Angular spectrum representation of a propagating wave}
An electromagnetic field $\mathbf{E}(x,y,z)$ in the space satisfies
the Maxwell equations. The electric field $\mathbf{E}(x,y,z\mathbf{)}$
can be expressed as
\begin{equation}
\mathbf{E}(x,y,z)=\stackrel[-\infty]{\infty}{\iint}\mathbf{\hat{E}}(k_{x},k_{y};z)e^{i(k_{x}x+k_{y}y)}dk_{x}dk_{y},\label{eq:E_field}
\end{equation}
where $\mathbf{\hat{\mathbf{E}}}(k_{x},k_{y};z)$ is the Fourier transform
of the electrical field on the $z={\rm constant}$ plane $\mathbf{E}(x,y,z)$:
\begin{equation}
\mathbf{\hat{\mathbf{E}}}(k_{x},k_{y};z)=\frac{1}{4\pi^{2}}\stackrel[-\infty]{\infty}{\iint}\mathbf{E}(x,y,z)e^{-i(k_{x}x+k_{y}y)}dxdy.\label{eq:E_field_spectrum}
\end{equation}
Considering the case that the medium in the space is source free,
linear, homogeneous and isotropic, the $\mathbf{E}(x,y,z)$ satisfy
the vector Helmholtz equation:
\begin{equation}
(\nabla^{2}+k^{2})\mathbf{E}(x,y,z)=0.\label{eq:Vector_Helmholtz_Eq}
\end{equation}
Substituting Eq.~(\ref{eq:E_field}) into Eq.~(\ref{eq:Vector_Helmholtz_Eq}),
we get the general solution (see similar materials on pp.~110
in Ref.~\cite{Mandel1995optical})
\begin{equation}
\mathbf{\hat{E}}(k_{x},k_{y};z)=\mathbf{\hat{A}}(k_{x},k_{y})e^{ik_{z}z}+\mathbf{\hat{B}}(k_{x},k_{y})e^{-ik_{z}z},\label{eq:E_field_spectrum_general}
\end{equation}
where we define $k_{z}=(k^{2}-k_{x}^{2}-k_{y}^{2})^{\frac{1}{2}}$
and ${\rm Im}(k_{z})>0$. On substituting Eq.~(\ref{eq:E_field_spectrum_general})
into Eq.~(\ref{eq:E_field}) the field
\begin{eqnarray}
\mathbf{E}(x,y,z) & = & \stackrel[-\infty]{\infty}{\iint}[\mathbf{\hat{A}}(k_{x},k_{y})e^{ik_{z}z}+\mathbf{\hat{B}}(k_{x},k_{y})e^{-ik_{z}z}]\nonumber \\
 &  & \times e^{i(k_{x}x+k_{y}y)}dk_{x}dk_{y}.\label{eq:E_field_full}
\end{eqnarray}
It is noted that so far we haven't specified the propagating direction
of $\mathbf{E}(x,y,z)$ or any other physical information. In fact,
$\mathbf{E}(x,y,z)$ expressed by Eq.~(\ref{eq:E_field_full})
has four parts: homogeneous wave propagating in the positive $\mathbf{z}$ direction
($e^{ik_{z}z}$, $k_{x}^{2}+k_{y}^{2}<k^{2}$), evanescent wave propagating
in the positive $\mathbf{z}$ direction ($e^{ik_{z}z}$, $k_{x}^{2}+k_{y}^{2}>k^{2}$),
homogeneous wave propagating in the negative $\mathbf{z}$ direction
($e^{-ik_{z}z}$, $k_{x}^{2}+k_{y}^{2}<k^{2}$), and evanescent
wave propagating in the negative $\mathbf{z}$ direction ($e^{-ik_{z}z}$,
$k_{x}^{2}+k_{y}^{2}>k^{2}$).

Now we consider an electromagnetic wave propagating along positive
$\mathbf{z}$ direction into the $z\geq z_{s}$ half space where $z_{s}\rightarrow-\infty$.
First, since we have known the direction of the wave, then $B(k_{x},k_{y})=0$.
Thus Eq.~(\ref{eq:E_field_spectrum_general}) is reduced
to 
\begin{equation}
\mathbf{\hat{E}}(k_{x},k_{y};z)=\mathbf{\hat{A}}(k_{x},k_{y})e^{ik_{z}z}.\label{eq:E_field_spectrum_positive}
\end{equation}
Also, setting $z=0$ in Eq.~(\ref{eq:E_field_spectrum_positive})
we get 
\begin{equation}
\hat{\mathbf{A}}(k_{x},k_{y})=\mathbf{\hat{E}}(k_{x},k_{y};0),
\end{equation}
where $\mathbf{\hat{\mathbf{E}}}(k_{x},k_{y};0)$ is the Fourier transform
of the electrical field on the $z=0$ plane $\mathbf{E}(x,y,0)$:
\begin{equation}
\mathbf{\hat{\mathbf{E}}}(k_{x},k_{y};0)=\frac{1}{4\pi^{2}}\stackrel[-\infty]{\infty}{\iint}\mathbf{E}(x,y,0)e^{-i(k_{x}x+k_{y}y)}dxdy.
\end{equation}
Second, $\mathbf{\hat{E}}(k_{x},k_{y};z_{s})=\mathbf{\hat{E}}(k_{x},k_{y};0)e^{ik_{z}z_{s}}$
with $z_{s}\rightarrow-\infty$ should be a finite value when $k_{x}^{2}+k_{y}^{2}>k^{2}$,
because it represents a physical field. Thus $\mathbf{\hat{E}}(k_{x},k_{y};0)=0$
when $k_{x}^{2}+k_{y}^{2}>k^{2}$. It means that a propagating wave
has none evanescent part (saying in another way, the evanescent wave
can't propagate along the positive $\mathbf{z}$ direction to far
from $z=z_{s}$). The electric field $\mathbf{E}(x,y,z\mathbf{)}$
in the $z\geq z_{s}$ half space with $|z-z_{s}|\gg0$ now reads
\begin{equation}
\mathbf{E}(x,y,z)=\underset{k_{x}^{2}+k_{y}^{2}\leq k^{2}}{\iint}\mathbf{\hat{E}}(k_{x},k_{y};0)e^{i(k_{x}x+k_{y}y+k_{z}z)}dk_{x}dk_{y}.\label{eq:E_field_positive}
\end{equation}

\subsection{Focusing of the propagating wave by aplanatic lens}

As shown in Fig.~\ref{fig:TheoreticalModel}, $\mathbf{E}_{{\rm inc}}(x,y,-f_{1})$
and $\mathbf{E}_{1}(x,y,-f_{1})$ are used to denote the field on
the left and right side of the lens $L_{1}$. $\mathbf{E}_{3}(x,y,f_{2})$
and $\mathbf{E}_{4}(x,y,f_{2})$ are used to denote the field on the
left and right side of the lens $L_{2}$. And
$\mathbf{E}_{2}(x,y,0)$ denotes the field on the $z=0$ plane. 

Using the stationary phase method \cite{CarterJOSA1972Elec,Mandel1995optical},
Eq.~(\ref{eq:E_field_positive}) can be evaluated.
For $z>0$ and $kr\rightarrow+\infty$, it reads
\begin{equation}
\mathbf{E}_{+\infty}(x,y,z)=-2\pi is_{z}k\mathbf{\hat{E}}(ks_{x},ks_{y};0)\frac{e^{ikr}}{r}.\label{eq:near-farfield-positiveDirec-zPositive}
\end{equation}
Thus substituting Eq.~(\ref{eq:near-farfield-positiveDirec-zPositive})
into Eq.~(\ref{eq:E_field_positive}) and using variable
substitution, it arrives
\begin{eqnarray}
\mathbf{E}(x,y,z) & = & \frac{ire^{-ifr}}{2\pi}\underset{k_{x}^{2}+k_{y}^{2}\leq k^{2}}{\iint}\mathbf{E}_{+\infty}(rt_{x},rt_{y},rt_{z})\nonumber \\
 &  & \times e^{i(k_{x}x+k_{y}y+k_{z}z)}\frac{1}{k_{z}}dk_{x}dk_{y}.\label{eq:E_field_positiveDirec_zpositive}
\end{eqnarray}
Here, we have defined those symbols: 
\begin{equation}
r=\sqrt{x^{2}+y^{2}+z^{2}},
\end{equation}
\begin{equation}
\mathbf{s}=(s_{x},s_{y},s_{z})=(\frac{x}{r},\frac{y}{r},\frac{z}{r}),
\end{equation}
\begin{equation}
\mathbf{t}=(t_{x},t_{y},t_{z})=(\frac{k_{x}}{k},\frac{k_{y}}{k},\frac{k_{z}}{k}).
\end{equation}
For $z<0$ and $kr\rightarrow-\infty$, using the same method, we
get
\begin{equation}
\mathbf{E}_{-\infty}(x,y,z)=-2\pi is_{z}k\mathbf{\hat{E}}(-ks_{x},-ks_{y};0)\frac{e^{-ikr}}{r},\label{eq:near-farfield-positiveDirec-zNegative}
\end{equation}
and 
\begin{eqnarray}
\mathbf{E}(x,y,z) & = & -\frac{ire^{+ifr}}{2\pi}\underset{k_{x}^{2}+k_{y}^{2}\leq k^{2}}{\iint}\mathbf{E}_{-\infty}(-rt_{x},-rt_{y},-rt_{z})\nonumber \\
 &  & \times e^{i(k_{x}x+k_{y}y+k_{z}z)}\frac{1}{k_{z}}dk_{x}dk_{y}.\label{eq:E_field_positiveDirec_zNegative}
\end{eqnarray}
It is noted that the focal point is usually chosen as the original
point of the axis frame as shown in Fig.~\ref{fig:TheoreticalModel}. The lens $L_1$ locates at $z=-f_{1}$,
so $kr\rightarrow-\infty$ is satisfied since $f_{1}$ is
much larger than the wavelength. Then Eq.~(\ref{eq:E_field_positiveDirec_zNegative})
describes the filed near the focus.

Usually the incident beam is linearly polarized Gaussian beam, so the field is with even symmetry. At this time, the field on the $z=-f_{1}$ plane
is even, i.e., 
\[
\mathbf{E}(-x,-y,z)=\mathbf{E}(x,y,z),
\]
then Eq.~(\ref{eq:E_field_positiveDirec_zNegative})
can be written as
\begin{eqnarray}
\mathbf{E}(x,y,z) & = & -\frac{ire^{ifr}}{2\pi}\underset{k_{x}^{2}+k_{y}^{2}\leq k^{2}}{\iint}\mathbf{E}_{-\infty}(rt_{x},rt_{y},-rt_{z})\nonumber \\
 &  & \times e^{i(k_{x}x+k_{y}y+k_{z}z)}\frac{1}{k_{z}}dk_{x}dk_{y}.\label{eq:E_field_positiveDirec_zNegative_even}
\end{eqnarray}
For linearly polarized LG beams, the vortex phase term $e^{il\phi}$ in $\mathbf{E}(x,y,z)$ will add a total phase $e^{il\pi}$ before the right side of Eq.~(\ref{eq:E_field_positiveDirec_zNegative_even}). Similar equations as Eq.~(\ref{eq:E_field_positiveDirec_zNegative_even}) can be found in Ref.~\cite{Novotny2012principles,JonesBook2015optical} (with the phase term in the front different, which can be omitted).

As a systematical formulation, we also list the relations between the focal field and
far field for the case of wave propagating along negative
$\mathbf{z}$ direction. For $z>0$ and $kr\rightarrow+\infty$:
\begin{equation}
\mathbf{E}_{+\infty}(x,y,z)=2\pi is_{z}k\mathbf{\hat{E}}(-ks_{x},-ks_{y};0)\frac{e^{-ikr}}{r},\label{eq:near-farfield-negativeDirec-zPositive}
\end{equation}
and 
\begin{eqnarray}
\mathbf{E}(x,y,z) & = & -\frac{ire^{ifr}}{2\pi}\underset{k_{x}^{2}+k_{y}^{2}\leq k^{2}}{\iint}\mathbf{E}_{+\infty}(-rt_{x},-rt_{y},rt_{z})\nonumber \\
 &  & \times e^{i(k_{x}x+k_{y}y-k_{z}z)}\frac{1}{k_{z}}dk_{x}dk_{y}.\label{eq:E_field_negativeDirec_zPositive}
\end{eqnarray}
For $z<0$ and $kr\rightarrow-\infty$:
\begin{equation}
\mathbf{E}_{-\infty}(x,y,z)=2\pi is_{z}k\mathbf{\hat{E}}(ks_{x},ks_{y};0)\frac{e^{ikr}}{r},\label{eq:near-farfield-negativeDirec-zNegative}
\end{equation}
and 
\begin{eqnarray}
\mathbf{E}(x,y,z) & = & \frac{ire^{-ifr}}{2\pi}\underset{k_{x}^{2}+k_{y}^{2}\leq k^{2}}{\iint}\mathbf{E}_{-\infty}(rt_{x},rt_{y},-rt_{z})\nonumber \\
 &  & \times e^{i(k_{x}x+k_{y}y-k_{z}z)}\frac{1}{k_{z}}dk_{x}dk_{y}.\label{eq:E_field_negativeDirec_zNegative}
\end{eqnarray}

\subsection{Relation of the field on the detector and the focusing lens}
Now we use the theory in last subsection to derive the relation between
$\mathbf{E}_{3}(x,y,-f_{2})$ and $\mathbf{E}_{1}(x,y,-f_{1})$. For
$\mathbf{E}_{1}(x,y,-f_{1})$, taking $-kf_{1}\rightarrow-\infty$,
according Eq.~(\ref{eq:near-farfield-positiveDirec-zNegative})
we get
\begin{equation}
\hat{\mathbf{E}}_{2}(k\frac{x_{1}}{f_{1}},k\frac{y_{1}}{f_{1}};0)=\frac{if_{1}e^{+ikf_{1}}}{2\pi s_{1z}k}\mathbf{E}_{1}(-x_{1},-y_{1},-f_{1}).\label{eq:near-far-3a}
\end{equation}
In aplanatic system, $r_{1}=f_{1}$ and $r_{2}=f_{2}$. For $\mathbf{E}_{3}(x,y,f_{2})$,
taking $kf_{2}\rightarrow+\infty$, according Eq.~(\ref{eq:near-farfield-positiveDirec-zPositive})
we get
\begin{equation}
\hat{\mathbf{E}}_{2}(k\frac{x_{2}}{f_{2}},k\frac{y_{2}}{f_{2}};0)=\frac{if_{2}e^{-ikf_{2}}}{2\pi s_{2z}k}\mathbf{E}_{3}(x_{2},y_{2},f_{2}).\label{eq:near-far-3b}
\end{equation}
Since the left sides of last two equations denote the same field,
using the variable substitutions $x_{1}=\frac{f_{1}}{f_{2}}x_{2}$
and $y_{1}=\frac{f_{1}}{f_{2}}y_{2}$ in Eq.~(\ref{eq:near-far-3a}),
the right sides of Eq.~(\ref{eq:near-far-3a}) and
Eq.~(\ref{eq:near-far-3b}) are equal. Then it arrives
\begin{equation}
\mathbf{E}_{3}(x_{2},y_{2},f_{2})=-e^{ik(f_{2}+f_{1})}e^{il\phi}\mathbf{E}_{1}(-\frac{x_{2}}{M},-\frac{y_{2}}{M},-f_{1}),
\end{equation}
where $M=\frac{f_{2}}{f_{1}}$ is the magnification factor. Usually
it is supposed that all the light ( s-polarized and p-polarized light
with various incident angle) transmits through the aplanatic lens
completely (transmission $t(p)=t(s)=1$), thus we get Eq.~(\ref{eq:Field_detector_incident}).

%\bibliographystyle{apsrev4-1}
%\bibliography{Bib_QPDSensitivity}

\end{document}